\documentclass{ppig}
\usepackage{epsfig}
\usepackage{booktabs}
\usepackage{ucs}
\usepackage[utf8x]{inputenc}

\usepackage[greek, english]{babel}
\usepackage[colorlinks,citecolor=teal, urlcolor=gray]{hyperref}
\usepackage{natbib}
\usepackage{todonotes}

\title{Intention Is All You Need}

\author{
  Advait Sarkar \\
  Microsoft Research, University of Cambridge, University College London \\
  advait@microsoft.com \\
}
\date{}

\begin{document}
\maketitle

% page numbers
\pagestyle{plain}

\begin{abstract}
Among the many narratives of the transformative power of Generative AI is one that sees in the world a latent nation of programmers who need to wield nothing but intentions and natural language to render their ideas in software. In this paper, this outlook is problematised in two ways. First, it is observed that generative AI is not a neutral vehicle of intention. Multiple recent studies paint a picture of the ``mechanised convergence'' phenomenon, namely, that generative AI has a homogenising effect on intention. Second, it is observed that the formation of intention itself is immensely challenging. Constraints, materiality, and resistance can offer paths to design metaphors for intentional tools. Finally, existentialist approaches to intention are discussed and possible implications for programming are proposed in the form of a speculative, illustrative set of intentional programming practices.
\end{abstract}

% A reflection. All human-computer interaction is programming. All programming is specifying behaviour. To interact with a computer through generative AI, you only need an intent, ostensibly. But AI supplies intent, too. It's mechanised convergence: automation making all work look the same. The vision of the programmer was one of control, but intent comes before control. So it turns out that programming, like human-computer interaction, is not really about specifying behaviour, but about intent. Intention is all you need. The existentialists got there first.

\section{The ``Intention Is All You Need'' Picture of Programming with Generative AI}
What is programming? Blackwell's succinct and influential definition is that programming is any activity exhibiting the property \emph{``that the user is not directly manipulating observable things, but specifying behaviour to occur at some future time''} \citep{blackwell2002programming}. Behaviour is specified through an interface, commonly a notation, which we call a programming language. Therein lies the source and objective of all research in the psychology and design of programming: the study of the use and improvement of the interfaces, notations, and languages for specifying behaviour.

The value of such study is called into question with the introduction of Generative Artificial Intelligence (\textbf{GenAI}), which can be defined as any \emph{``end-user tool [...] whose technical implementation includes a generative model based on deep learning''}.\footnote{There is no definitional consensus on this term. A rationale for the definition adopted here is given by \cite{sarkar2023eupgenai}.} GenAI captures the relationships between natural language specifications of behaviour, and the translations of that behaviour into programming notation, implicit in enormous training datasets. The power of translation thus captured can be stochastically replayed on demand \citep{blackwell2020objective}. What could this mean for research in the user-centred design of programming languages? One perspective anticipates nothing less than its obsolescence:

\begin{quotation}
\emph{``The programming barrier [with GenAI] is incredibly low. We have closed the digital divide. Everyone is a programmer now - you just have to say something to the computer''}\footnote{Huang, 2023. (Source: \url{https://www.reuters.com/technology/ai-means-everyone-can-now-be-programmer-nvidia-chief-says-2023-05-29/}, accessed July 2024.)}
\end{quotation}

% \begin{flushright}
% -- Jensen Huang, Nvidia CEO, May 2023\footnote{\url{https://www.reuters.com/technology/ai-means-everyone-can-now-be-programmer-nvidia-chief-says-2023-05-29/}, accessed July 2024}
% \end{flushright}

\begin{quotation}
\emph{``Up until now, in order to create software, you had to be a professional software developer. You had to understand, speak and interpret the highly complex, sometimes nonsensical language of a machine that we call code. [... But with GenAI] We have struck a new fusion between the language of a human and a machine. With Copilot, any person can now build software in any human language with a single written prompt. [...] going forward, every person, no matter what language they speak, will also have the power to speak machine. Any human language is now the only skill that you need to start computer programming.''}\footnote{Dohmke, 2024. (Source: \url{https://www.ted.com/talks/thomas_dohmke_with_ai_anyone_can_be_a_coder_now}, accessed July 2024.)}
\end{quotation}

% \begin{flushright}
% -- Thomas Dohmke, GitHub CEO, April 2024\footnote{\url{https://www.ted.com/talks/thomas_dohmke_with_ai_anyone_can_be_a_coder_now}, accessed July 2024} 
% \end{flushright}

\begin{quotation}
\emph{``Since the launch of GPT-4 in 2023, the generation of whole apps from simple natural language requirements has become an active research area. [...] Our vision is that by 2030 end users will build and deploy whole apps just from natural requirements.''}\footnote{\citet{robinson2024requirements}}
\end{quotation}

\begin{quotation}
\emph{``Programming will be obsolete. [...] the conventional idea of `writing a program' is headed for extinction [...] all programs in the future will ultimately be written by AIs, with humans relegated to, at best, a supervisory role. [...] The engineers of the future will, in a few keystrokes, fire up an instance of a four-quintillion-parameter model that already encodes the full extent of human knowledge (and then some), ready to be given any task required of the machine.''}\footnote{\citet{welsh2022endofprogramming}}
\end{quotation}

% ``Programming will be obsolete. I believe the conventional idea of “writing a program” is headed for extinction, and indeed, for all but very specialized applications, most software, as we know it, will be replaced by AI systems that are trained rather than programmed. In situations where one needs a “simple” program (after all, not everything should require a model of hundreds of billions of parameters running on a cluster of GPUs), those programs will, themselves, be generated by an AI rather than coded by hand. [...] It seems totally obvious to me that of course all programs in the future will ultimately be written by AIs, with humans relegated to, at best, a supervisory role. [...] The engineers of the future will, in a few keystrokes, fire up an instance of a four-quintillion-parameter model that already encodes the full extent of human knowledge (and then some), ready to be given any task required of the machine.''

% \todo[inline]{\url{https://www.ted.com/talks/thomas_dohmke_with_ai_anyone_can_be_a_coder_now}
% \url{https://www.reuters.com/technology/ai-means-everyone-can-now-be-programmer-nvidia-chief-says-2023-05-29/}
% Adg - requirements are all you need
% Kind of like what Ben Rosen said about spreadsheets in the 1980s
% Scratch - Programming for all}

The promise of GenAI for programming, therefore, is to transform programming into an activity where expertise in specialised notations and languages for specifying behaviour are unnecessary. One merely has to say what one wishes the program to do, and GenAI does the rest. The interaction design challenges of programming are solved.\footnote{One is reminded of similar claims made during the early days of spreadsheets or about any number of visual programming languages. E.g., Benjamin Rosen, a PC industry analyst for Morgan Stanley, later a key funder of Lotus and Compaq, noted in 1979 that \emph{``In minutes, people who have never used a computer are writing and using programs [...] the user need not know \underline{anything} about computers or programming in order to derive Visicalc's benefits. You construct a Visicalc program much as you would define a problem on a sheet of paper or a blackboard''} \citep{rosen1979visicalc}.} Intention is all you need.

% I'm being facetious, of course. 

There are many problems with this picture. There are compelling reasons for continuing to engage with formal notations, even and perhaps especially when GenAI is in play \citep{sarkar2023eupgenai}. Moreover, language in general, and the language of prompts used to direct GenAI in particular, is most certainly not a flawless, transparent route for the expression of intent. Johnny can't prompt \citep{zamfirescu-pereira2023johnny}. Johnny can't figure out what level of abstraction to write his prompts in, either \citep{liu2023gam,sarkar2022programmingai}. Thinking about prompting is hard for Johnny, and thinking about thinking about prompting is hard, too \citep{tankelevitch2024metacognitive}. Prompting ``dialects'' might evolve in much the same way as specialised uses of natural language do in domains such as scientific and legal communication, through disciplinary norms and professional consensus, and to acquire such language will require users to undergo analogous processes of disciplinary and professional acculturation \citep{sarkar2023eupgenai}. But these problems are not the primary concern in this paper.

There is a rather more fundamental pair of problems with the idea that intention is all you need (to program with GenAI): it assumes that GenAI does not interfere with intention. Moreover, it takes for granted that intentions are easy to form. Both premises will be questioned in turn.

\section{Mechanised Convergence: The Homogenising Effect of AI on Intention}

Contrary to not interfering with intention, AI supplies intention. It does so in a way that can be described as \emph{mechanised convergence} \citep{sarkar2023creativity}, drawing on Walter Benjamin's concept of mechanical reproduction \citep{benjamin1935work}. Mechanised convergence describes the idea that the automation or mechanisation of work leads to a convergence in the space of outputs. Standardisation is necessary for factory logic to function. For a machine to be repeatable at speed, its inputs and outputs need to be repeatable at speed, too. You can have any colour as long as it's black.

% Benjamin noted how mechanical reproduction enabled formerly unique works of art to be experienced detached, en masse \citep{benjamin}.

Here is some of the evidence that GenAI has a mechanised convergence effect:

% - intentionality and AI; goal-shaping, most knowledge work has weak intentions, predictive text predictable writing, mechanised convergence - Walter Benjamin, Ivan illich, McLuhan.
% - technological reversal - section 5.3 of chiwork paper, also mention convergence in the BCG study: \url{https://papers.ssrn.com/sol3/papers.cfm?abstract_id=4573321}
% - see some other examples of papers about mechanised convergence cited in Michael's VL '24 submission
% - if Michael's VL '24 submission accepted, cite that too
% - discussion around 'satisficing' from EuSpRIG paper

% Nikhil Sharma, Q. Vera Liao, and Ziang Xiao. 2024. Generative Echo Chamber? Effect of LLM-Powered Search Systems on Diverse Information Seeking. In Proceedings of the CHI Conference on Human Factors in Computing Systems (CHI '24). Association for Computing Machinery, New York, NY, USA, Article 1033, 1–17. https://doi.org/10.1145/3613904.3642459
\begin{itemize}
\item Predictive text encourages predictable writing \citep{arnold2020predictive}. In an image captioning task, when participants use predictive text entry systems, captions written with suggestions are shorter and use fewer words that the system does not predict. A similar effect occurs in identifier names when programmers use a GenAI tool such as GitHub Copilot to assist them in writing code \citep{lee2024predictability}. This effect occurs even when the suggestions are merely visible and not actionable (i.e., cannot be accepted using a keyboard shortcut).

\item Similarly, a large study (n=293) of participants writing short stories with varying degrees of AI assistance found that exposure to GenAI ``ideas'' leads to a reduced diversity of content \citep{Doshi2023GenerativeAI}. Participants exposed to even a single GenAI suggestion produce stories similar to the average of the other stories in the same experimental condition.

\item A large study (n=758) of strategy consultants at BCG examined the effects of ChatGPT use on a set of consultancy tasks \citep{dell2023navigating}. The majority of participants with access to ChatGPT retain a very high amount of its response -- typically around 90\% -- in their submitted work. Participants without access to ChatGPT produce ideas with more conceptual variation than those with access, showing that usage of ChatGPT reduces the range of ideas generated. The variation across responses produced by ChatGPT is smaller than what human participants produce on their own.

\item Large language models have a ``homogenization effect'' on creative ideation \citep{anderson2024homogenization}. In a creative ideation task, participants produce less semantically distinct ideas when using ChatGPT. Moreover, participants feel less responsible for ideas produced with ChatGPT assistance.

\item A large study (n=115) finds that conversational search built on GenAI increases selective exposure compared to conventional search \citep{sharma2024echo}. Users engage in more biased information querying with conversational search, and the bias is exacerbated when the model is itself ``opinionated'' to reinforce the user's views. The authors call this a ``generative echo chamber''. 

\item Similarly, a large study (n=1506) of co-writing with GenAI found that using an ``opinionated'' language model affects the opinions expressed in participants' writing and moreover, actually shifts their opinions as measured in a subsequent attitude survey \citep{jakesch2023opinionated}. A related effect, termed ``drifting'', has been observed in novice programmers, where the tendency to accept and adapt code generated by the system leads programmers away from a correct solution \citep{prather2024drifting}.
\end{itemize}

Mechanised convergence signals an odd reversal (or perhaps intensification) of Dennett's ``intentional stance'' \citep{dennett1971intentional}, wherein we not only ascribe intention to these systems but also delegate it, sometimes wilfully, other times unknowingly.

The intention supplied by GenAI through mechanised convergence has a complex source, combining influences of its training data, and the biases and heuristics encoded by the system developers. However at its core, mechanised convergence is the ultimate outcome of the old statistical logics of uncovering underlying natural ``laws'' \citep{blackwell2020objective,sarkar2023humanaicollaboration}. The statistical machine eliminates ``noise'' (diversity) to predict ``signal'' (uniformity). The statistical machine is the triumph of the Enlightenment aesthetic faith in nature's having an underlying elegance or simplicity that is obscured from view by imperfect forms. It should come as no surprise that machines that are built to search for Platonic ideals reflect back to us a mechanically converged picture of the world, making quiddity of haecceity.
% , regardless of whether such a world exists.

It is important to note that the effect on intent as demonstrated in these studies is an \emph{aggregate tendency} that likely does not square with individual phenomenal perceptions of GenAI use. At the granularity of individual interactions, the experience of GenAI might well be as a passive translator, not active supplier, of intent. The nudge towards standardised, centralised, averaged, generic, and statistically optimised answers may be barely perceptible. Yet the data demonstrates that these nudges in fact have a measurable cumulative effect on knowledge work.

% Langdon: convergence is the politics of AI, the artefact.
As Winner sets out, artefacts have politics \citep{winner1980artifacts}. The design features of a technology enable certain forms of power, and the decision to adopt a particular technology requires certain power relations to be enacted. Putting it in Winner's terms, convergence is the politics of AI, the artefact.

% McLuhan: convergence is the message of AI, the medium.
As McLuhan sets out, the medium is the message \citep{mcluhan1964media}. There is an effect of a particular medium, be it typography, radio, or television, on the human sensorium that is quite distinct from any particular content being conveyed through that medium. The effect of the medium overwhelms the content and makes it incidental. Putting it in McLuhan's terms, convergence is the message of AI, the medium.

McLuhan predicted that electric technology and programmability would reverse the convergence tendencies of factory logic. He gives the example of a programmable tailpipe machine: \emph{``A new automatic machine for making automobile tailpipes [...] starting with lengths of ordinary pipe, it is possible to make eighty different kinds of tailpipe in succession, as rapidly, as easily, and as cheaply as it is to make eighty of the same kind. And the characteristic of electric automation is all in this direction of return to the general-purpose handicraft flexibility that our own hands possess. The programming can now include endless changes of program.''} 

Taken to its logical conclusion, McLuhan makes a claim that is strikingly similar to the narrative that intention is all you need: \emph{``the older mechanistic idea of “jobs,” or fragmented tasks and specialist slots for “workers,” becomes meaningless under automation. [...] The very toil of man now becomes a kind of enlightenment. As unfallen Adam in the Garden of Eden was appointed the task of the contemplation and naming of creatures, so with automation. We have now only to name and program a process or a product in order for it to be accomplished. Is it not rather like the case of Al Capp’s Schmoos? One had only to look at a Schmoo and think longingly of pork chops or caviar, and the Schmoo ecstatically transformed itself into the object of desire. Automation brings us into the world of the Schmoo. The custom-built supplants the mass-produced.''} As we have seen, the vast programmability of GenAI does not necessarily result in a \emph{``return to [...] general-purpose handicraft flexibility''}, rather, it has enabled a newer, subtler, and more pervasive form of the \emph{``fragmentalized and repetitive routines of the mechanical era''}. Through the mechanised convergence of knowledge work through GenAI, the principle of interface design becomes WYGIWYG -- What You Get Is What You Get.

Postman, who builds on McLuhan, more accurately reappraised the effect of the electric age on intention \citep{Postman1985-rv}. He explains that the information age has resulted not in an Orwellian dystopia where intentions are surveilled and constrained, but rather a Huxleyan one, where intentions are numbed: \emph{``What Orwell feared were those who would ban books. What Huxley feared was that there would be no reason to ban a book, for there would be no one who wanted to read one. Orwell feared those who would deprive us of information. Huxley feared those who would give us so much that we would be reduced to passivity and egoism. Orwell feared that the truth would be concealed from us. Huxley feared the truth would be drowned in a sea of irrelevance. Orwell feared we would become a captive culture. Huxley feared we would become a trivial culture [...]''}. We inhabit not Foucault's society of discipline \citep{Foucault1977-hi, o1986disciplinary}, but Deleuze's society of control \citep{deleuze1992control}. 

This scenario is undesirable, not least because mechanised convergence implies a reduction in the rate at which new ideas are generated, and an increase in repetition and replay of existing ideas. What kind of culture springs from the consumption and emission of an increasingly convergent set of increasingly recycled ideas? A derivative, ``stuck'' culture, is the diagnosis of technology critic Paul Skallas.\footnote{\url{https://lindynewsletter.beehiiv.com/p/culture-stuck}, accessed July 2024. Related is the concept of ``refinement culture''; \emph{``Refinement culture can be summarized as a general streamlining and removal of any unique characteristics. Refinement Culture is one dimensional and removes variety from the environment. It's optimized.''} \url{https://lindynewsletter.beehiiv.com/p/refinement-culture}, accessed July 2024, \url{https://medium.com/@lindynewsletter/refinement-culture-51d96726c642}, accessed 2024.} Even for GenAI itself, the indications are that the roads of autophagy lead to madness; the roads of recursion lead to cursed collapse \citep{alemohammad2023selfconsuming, shumailov2024collapse, bohacek2023nepotistic, gerstgrasser2024recursion}.

% \todo[inline]{Intention - stuck culture, refinement culture}

% [...] In 1984, Huxley added, people are controlled by inflicting pain. In Brave New World, they are controlled by inflicting pleasure. In short, Orwell feared that what we hate will ruin us. Huxley feared that what we love will ruin us.''

Mechanised convergence, as a tendency of automation more broadly, creates a crisis of intentionality: a culture that has lost the capacity to intend, does not realise, and does not care.

% Crisis is deeper than agency. Perception of agency can be manipulated but can be false. Intention can be manipulated too but this is a true modification of intention.

% \section{Intention}

\section{Interlude: Babbage's Intentional Programmer}
Describing what GenAI does to intention as a ``crisis'' implies that we need to do something about it. Indeed, what we need to do about it is to promote the active cultivation of the capacity to intend.\footnote{Much as \cite{benjamin2024imagination} calls for us to cultivate the capacity to imagine.}

Since this is PPIG, we can start by considering the intentions of programmers. What the tendency for mechanised convergence tells us is that, prior to specifying behaviour, programming must be about forming an intention for behaviour. A definition of programming that centres intention, rather than specification, evokes a rather older philosophy of programming that we can draw from the crisis in theology at the time of Babbage.

% An older philosophy of programming, from Babbage.

Science (more precisely, natural philosophy) in post-Enlightenment Britain at the time of Babbage was grappling with the apparent contradiction of divine miracles -- acts of God outside the laws of nature created by God -- which Hume had famously argued could not be rationally supported \citep{hume1748enquiry}. In aiming to discover mathematical laws such as those of Newton, which could accurately describe and predict nature, natural philosophers operating within the frameworks of Deism and Christianity struggled to reconcile their work and faith.

Babbage found in his Difference Engine the possibility for reinterpreting miracles as part of the natural divine order. Using a ``feedback mechanism'' that connected two gear wheels, Babbage was able to encode programs that, after a certain number of iterations, would change their behaviour. For example, he would demonstrate a program that counts the integers 1, 2, 3 ... up to 100, at which point the program would change and start counting in steps of two: 102, 104, 106 ... etc. In demonstration-sermons delivered to rapturous audiences, he used this example to explain his theory of God as a \emph{divine programmer} \citep{Snyder2012-yt}. A miracle was thus explained as a shift in the program. God's intervention to perform apparent miracles was not an aberration against universal, constant laws -- it was merely the manifestation of a deeper and misunderstood universal law, a deeper plan, a deeper intention.

It is instructive that Babbage's conception of programming and intention centred around shifts, or deviations, from the expected. A machine that continues to execute the same predictable behaviour is not a program, it is simply a machine. It is in the departure from convergent behaviour that evidence of programming emerges as activity and divinity. For Babbage, to converge is human, to deviate divine. To execute is human, to program divine. To specify is human, to intend divine.

\section{Sources of Intention: Constraints, Materiality, and Resistance}
% Where can we get intentions from? Evolutionary/psychological basis not enough. Free will and neuroscience.
% Where do intentions come from? 

Returning to our objective -- to promote the active cultivation of the capacity to intend -- it is worth briefly exploring a few perspectives on the sources of intentions.

% Intentions in response to external world. Creativity and constraints. Craftsmanship and materiality.
Much intention appears to arise as a result of interaction with the external world. Practitioners of creative arts and research in creativity have long noted the role of constraints in shaping and facilitating creativity \citep{stokes2005creativity,May1975-vg}. Materiality and resistance are essential to craftsmanship; any material, by virtue of its properties and resistances, participates in an ongoing dialogue with the craftsman's intentions \citep{basman2016building}. According to material engagement theory\footnote{Thanks to Ava Scott for identifying this connection.} \citep{malafouris2019mind}, \emph{``Our forms of bodily extension and material engagement are not simply external markers of a distinctive human mental architecture. Rather, they actively and meaningfully participate in the process we call mind''}. As such, the role of material as a source of intention can be seen as a form of extended cognition, or at the very least external cognition \citep{turner2016distributed}, notwithstanding challenges to these ideas \citep{rupert2004challenges}.

A sculptor must consider how pliable or fragile their material is, what tolerances and fine details can be accomplished, how gravity will constrain the scale and orientation of their figures. A carpenter must consider the grain of their wood, where cuts and incisions can be made. A painter using watercolours must consider and exploit the additive translucency of that medium, one using oils must consider the opacity of theirs. It is telling that the archetypical dimension in the Cognitive Dimensions of Notations \citep{green1989cognitive} is \emph{viscosity}, a metaphor rooted in materials and resistances, aiming to bridge them with the seemingly immaterial and disembodied world of notations.

% \todo[inline]{\url{https://link.springer.com/article/10.1007/s11097-018-9606-7}}

Some intentions even rejoice in the contradiction of others: for example, the objective of subversive gameplay styles is to ignore the received goals of the game and invent one's own \citep{Flanagan2009-hf}, it is playing the infinite game whose objective is to continue playing, not the finite game whose objective is to win \citep{Carse1986-fs}. Solving the continuous puzzles posed by these resistances, having a vision pushed, pulled, and evolved, is the pleasure and intentionality of craftsmanship. These are not destructive resistances that hinder the realisation of an intention; they are productive ones that facilitate it. 

% Exploratory programming, problem formulation as problem solving, interactive machine learning, constructivist design.
Exploratory programming \citep{kery2017exploring} exemplifies how the materialities and resistances of programming are exploited to shape intention. In exploratory programming, the programmer's goal is unknown or ill-defined. The objective of the process is to discover or create an intention, to formulate a problem. The formulation of a problem co-exists with and cannot be separated from its solution \citep{rittel1973dilemmas,sarkar2023simplicity}. This is also the case in the end-user programming activity of interactive machine learning, or interactive analytical modelling \citep{sarkar2016phd}, where the goal is ill-defined and the objective is to create one, through a constructivist loop of interaction between ideas and experiences \citep{sarkar2016constructivist}.

% AI as critic or provocateur, antagonistic AI, cognitive glitches, etc.
There have been proposals to design GenAI systems that introduce productive resistances as catalysts for the development of intention. Rather than an assistant, AI can act as a critic or provocateur \citep{sarkar2024challenge,Sarkar2024eusprig}. AI can be antagonistic \citep{cai2024antagonistic}. AI can cause cognitive glitches \citep{hollanek2019non}. AI can act as cognitive forcing functions \citep{bucinca2021trust}. These proposals are counter to traditional narratives of system support, system disappearance, and system non-interference. They can be seen as successors to previous counternarratives raised by researchers such as critiques of the doctrines of simplicity and gradualism \citep{sarkar2023simplicity}, critiques of seamlessness \citep{chalmers2003seamful}, critiques of reversible interactions \citep{rossmy2023point}, the case for design frictions and microboundaries \citep{cox2016design}, reframing of ambiguity as design resource \citep{gaver2003ambiguity}, and calls for attention checks in AI use \citep{gould2024chattldr}.\footnote{It is worth observing that while such counternarratives often involve calls for greater, more critical, and more reflective user engagement and participation with technology, it should not be assumed that intentionality always entails participation or action. Observation itself is not intrinsically passive. This point is well made by \cite{pfaller2017interpassivity}: \emph{``Two philosophical premises silently played a decisive role in this triumphal march of participation: first, the idea that the relation between transmitter and receiver represents a hierarchy and that the elimination of this hierarchy therefore has to amount to a democratisation; and secondly, the idea that it is more desirable for spectators to participate than to spectate [... however,] the relation between transmitter and receiver does not always represent a hierarchy. And when it does, then it is not always in favour of the transmitter [...] This is why it is misleading to believe that activating the audience in art is automatically and always tantamount to their liberation. Because could not the exact opposite be the case: could the enthusiasm for joining in produced by art not deprive people of the necessary refractoriness that they would need in political life in order not to be immediately enthused by every neoliberal or reactionary or even fascist appeal to ‘actively’ join in, and pursue this with a feeling of liberation?''}}

% \todo[inline]{
% % Design approaches to intentionality: critical thinking designs, reflective computing, forcing functions, simplicity, seamful computing, ambiguity as a resource for design, point of no undo, chattldr attention checks. Intentionality vs participation. Observation is not passive (Zizek, Pfaller interpassivity). Also Intention - Illich vernacular? Design frictions for mindful interactions: The case for microboundaries (Cox et al.). \url{https://www.interaction-design.org/literature/book/the-glossary-of-human-computer-interaction/forcing-functions}. 
% Intentional stance - Dennett. Kewenig intentionality and consciousness in robot love. Intentionality vs participation. Observation is not passive (Zizek, Pfaller interpassivity). Also Intention - Illich vernacular?}

% Resistance is the opposite of support.

The concept of resistance could be key to framing the design objectives for intentional GenAI tools. Our current explorations of improving critical thinking with GenAI (e.g., \citet{Sarkar2024eusprig}) are strictly \emph{additive}: let's augment AI interaction and output with prompts, text, visualisations, etc. that get the user thinking. However, this approach increases the cognitive burden by asking users to consume and reflect on more information. We know that people don't always enjoy, or want, more information. Particularly when it comes to the user interfaces of discretionary software, they usually want less \citep{carroll1987paradox,sarkar2023simplicity}. The additive approach may be starting by fighting a losing battle, one in which we try to design the smallest, most stimulating, most rewarding ``consumable'' that creates user reflection, without incurring undesirable attentional costs. The idea of resistance provides a different starting point. How can we build GenAI tools with inherent, productive resistances that are part of working with the tool, not an additional thing that users need to ``pay'' attention to? How can the experience of resistances in the interface feel more like the pliability of clay, or the translucency of paint? This is an open avenue for future work.

% Existentialist philosophy. End with something about how it teaches us that we are in a process of becoming.

\section{Existentialist Approaches to Intention}
So far we have been considering intention at relatively small scale: instances of knowledge work and GenAI use. But intentions, like goals, form hierarchies. Intentions are not isolated and independent, they are related and convergent. To what do they converge? At this point we shall make a somewhat abrupt leap outwards and consider the most expansive scope of intention -- as enacted over the course of an entire life. 

An evolutionary account might attempt to trace human intentions back to fundamental physiological concerns: we form intentions to continue survival, to avoid fear, to ensure comfort, to maximise pleasure, to minimise pain. These can certainly account for some intentions. The concept of intention has much in common with free will -- loosely defined, one's capacity to act differently to how one did, in fact, act. Free will is not the same as intention, but it can be viewed as a precondition for true intention. Neuroscientific work purporting to demonstrate (a lack of) free will has been criticised by philosophers because (among other objections), we do not have a suitably good picture that connects short-term choices dominated by low-level psychological phenomena (such as choosing to push the left button or the right button) to the complex, long-term, highly planned and goal-oriented intentions (such as the intention to commit a crime) that pose the truly consequential ethical challenges to free will \citep{mele2019free}. The evolutionary account is part of a broader category of \emph{teleosemantic} theories of intention \citep{sep-intentionality} according to which design (evolutionary or artificial) supplies a function (\textgreek{τέλος}), which in turn supplies intention.

% Payal Arora Maslow's hierarchy reverse. If not reverse then at least tradeoff.
In considering whether human intention can truly be reduced to evolutionary or functional needs, I am drawn to the argument made by feminist anthropologist Payal Arora in her closing keynote for the 2022 CHI conference \citep{Arora2022}. She criticizes Maslow's famous hierarchy of needs that places physiological and safety needs at the bottom, rising to esteem and self-actualisation at the top. The conventional reading is that needs at the bottom of the pyramid need to be satisfied, the foundation of the pyramid needs to be built, before one can proceed to the higher levels. This is a fairly influential way of thinking and often dictates the way in which social aid and rescue efforts are prioritised: focus on food, water, and shelter first, and joy, play, growth, education, and dignity later. Arora finds that this picture does not correspond with her observations in her extensive ethnographic work with precarious, oppressed, and underprivileged groups. Instead, she proposes that the pyramid is upside down. What she finds is that self-actualisation is what people need first, and are willing to sacrifice safety needs to get it. People leave secure work when the nature of that work threatens their dignity, even if this places them in financial hardship. People leave homes where they cannot express their identity, or are not accepted for who they are, even if this might leave them without a roof over their head. A line from the poet James Oppenheim captures the sentiment: 
\begin{quote}
    \emph{``Our days shall not be sweated from birth until life closes ---\\ 
    Hearts starve as well as bodies: Give us Bread, but give us Roses.''} 
\end{quote}

If not entirely upside down, then at the very least Maslow's hierarchy is not a unidirectional ladder to climb, but a set of considerations and influences that are continually negotiated and traded-off. Physiology and evolution are part of intention formation, but far from the entire picture. Where can we look for a perspective on intention that aligns with Arora's observations? Moreover, is there an approach that not only identifies the source of intention, but prescribes a method for cultivating it?

Elaborating the consequences of the idea that the active cultivation of intention is \emph{the} core virtue in an inherently meaningless world is precisely the project of existentialist philosophy. 

The absence of any inherent purpose to life is the starting point. Per \cite{Sartre1956}, ``existence precedes essence''; individuals first exist without purpose and must subsequently forge their essence, or identity, through their actions. Angst, or existential anxiety, arises from the realization of one's freedom and the infinite possibilities it entails \citep{Kierkegaard1844}. Existentialists see angst as a motivator rather than an obstacle.

Authenticity is one expression of existentialist intention. It is the pursuit of living in accordance with one's true self and values, rather than conforming to societal norms, and is essential for genuine existence \citep{heidegger1927being}. Authenticity requires a conscious effort to understand and act upon personal convictions, even in the face of adversity or societal pressure \citep{kierkegaard1843fear, deBeauvoir1948}. Other sources of intentionality, besides authenticity, go beyond the individual. Kierkegaard's \citep{kierkegaard1849sickness} ``leap of faith'' suggests that to escape from existential despair requires acknowledging the limits of rational reflection and an individual's relationship with the divine. Moreover, to seek engagement with the world is to step beyond oneself, to interact with others, and to find and create meaning through these actions \citep{jaspers1932philosophie}. Similarly, \citet{deBeauvoir1948} points out that our individual subject-like freedom is complemented by an object-like unfreedom (``facticity''), deriving an ethics of freedom that advocates for actions that respect the freedom of others.

\cite{Camus1955} counsels individuals to accept ``the absurd'': the tension between the human search for meaning and a universe that is silent in response, to recognize the lack of inherent meaning in the world and to take on the task of creating their own purpose. Camus rejects ``solutions'' to the absurd proposed by prior philosophers, such as Kierkegaard, as ``philosophical suicide''. To Camus, seeking overarching meaning despite the absurd is seeking to resolve, minimise, sidestep, or ignore the absurd, not acknowledging it. 

Camus rejects a forced imposition of meaning where there is none. A leap of faith is a form of escape. Incidentally, a forced imposition of meaning is precisely the \emph{modus operandi} of GenAI: for language to be produced by arithmetic means it is necessary to encode language in a uniform, rational vector space. Sense and nonsense alike are thus enumerated and made commensurable. \emph{King$-$Man$+$Woman$=$Queen} \citep{mikolov2013linguistic}. Before carefully designed guardrails (themselves a form of escape) made it more difficult to do so, it was easy to elicit answers to nonsense questions such as ``what colourless green ideas sleep furiously?'' from language models. Furthermore, GenAI is an essential component of an emerging pseudoreligious meta-narrative of escape identified by \cite{gebru2024tescreal}: \emph{\textquotedblleft What ideologies are driving the race to attempt to build AGI? [...] we trace this goal back to the Anglo-American eugenics movement, via
transhumanism. [...] we delineate a genealogy of interconnected and overlapping ideologies that
we dub the `TESCREAL bundle,' where the acronym `TESCREAL' denotes `transhumanism,
Extropianism, singularitarianism, (modern) cosmism, Rationalism, Effective Altruism, and longtermism'\textquotedblright}.  

% \todo{(Intention genai meaning like religion - tescreal bundle)}

% Existentialist philosophy posits that human beings are free agents who have control over their actions and decisions.

% The concept of the absurd \citep{Camus1955} further complicates human existence by highlighting the inherent meaninglessness of life and the struggle to find or create meaning within it. 

% This acceptance is not a resignation to nihilism but an invitation to actively engage in the construction of meaning through one's actions and choices.

% \todo[inline]{Elaborate on absurd, philosophical suicide is denying the absurd and resolving it by seeking meaning anyway, this is what GenAI does - the forced imposition of meaning}

Camus' existentialist view offers a non-escapist alternative that stares meaninglessness in the face and from it derives freedom. This freedom is both liberating and burdensome. We are at liberty to choose, but are also responsible for bearing the burden of the consequences. The lightness of being can thus be unbearable. It is through confronting this anxiety that individuals can make deliberate and meaningful choices, shaping their intentions, and by extension, their essence.

GenAI has implications for the intention of professional programmers and casual ones alike. The introduction poses the question ``what is programming?'', and we can now see a second reading of this question which asks not for a definition of an activity, but of an aspiration or identity. As GenAI solves the problem of control, of specifying behaviour, the aspiration shifts to intent. Intent precedes control. To be a programmer is therefore not to be one who specifies behaviour, but one who forms authentic, meaningful intentions for behaviour.

\section{Speculative Scenarios for Intentional Programming}

The optimism of the ``intention is all you need'' narrative does posit a legitimate observation concerning the behavioural economics of software production. GenAI makes the production of bespoke software vastly cheaper. One can view existentialism as a response to the loss of the ``grand narratives'' of modernity. But software has still been constrained by the grand narratives of capitalism and utility -- until now. To write a program required \emph{investment} of time and hard-earned expertise, exerting pressure on programs to be valuable, robust, and reusable. Where they did not place an outright barrier, the investment costs of programming disincentivised exploration, error, and disposal. Within this frame story hitherto sits the universe of programmer psychology and behaviours: from authoring code to code comprehension, from knowledge sharing and documentation to debugging, from learning barriers to attention investment, from API design to autocomplete. Almost the entire diversity of experience of programmers, professional or casual, that our research community has so carefully documented and explained for the last half-century, has dwelt in the shadow of the market's invisible hand.

As the hand is withdrawn, one might ask how programmers can respond, in a microcosm of the existential dilemma, to the liberating yet burdensome freedom granted by GenAI. As far as practical advice (i.e., ``implications for design[ing your life]'') is concerned, existentialists advise embracing one's freedom to shape life, living authentically, accepting the absurd, confronting anxiety, and seeking engagement with the world as ways to form meaningful intentions. What this might mean for programmers, and interaction with GenAI, can be sketched in a few speculative scenarios:

\begin{itemize}
    \item Intentional coding retreats: The programmer steps away from her standard way of working to participate in an intentional coding retreat. Here, the programmer reconnects with the craft of coding without the assistance of AI tools. This allows the programmer to explore and reaffirm personal coding styles and problem-solving approaches. For example, a programmer accustomed to relying on AI for debugging might rediscover the satisfaction of manually untangling complex code, thus reaffirming their individual capability and creative freedom.

    \item AI as muse: AI suggests an unusual, contradictory, or incorrect algorithmic approach, which the programmer then refines and transforms with personal insights and expertise. The tool is not a crutch but a source of inspiration.

    % \item Value-based generation: GenAI tools could require users to articulate their core values before generating outputs. For instance, a tool might prompt users to state values such as creativity, diversity, or integrity, and these choices would directly influence the behaviour of the system.

    \item Programming with provocations: programming environments include prompts or questions to stimulate deeper thinking about the purpose and potential impact of the code being written. This can help programmers reconnect with their motivations and aspirations.

    \item Programming with constraints: intentional constraints are introduced to programming projects, much like the practices of constrained writing.\footnote{E.g., see discussion of conceptual writing in \cite{sarkar2023creativity}.} Programmers already practice genres of constrained programming for pleasure, such as ``code golf'' (writing the shortest possible program with a certain behaviour) or ``quines'' (inputless programs that produce only their own source code as output). By deliberately limiting certain resources or imposing unique challenges, programmers can stimulate creativity and craft intentional solutions.

    \item Deviation practice: in the education of professional programmers, exercises are developed that require intentional deviation from established patterns. By practising the precise skill of breaking away from standard solutions, programmers may more readily acquire the conscious muscle and desire for forming unique intentions and exploring novel paths.

    \item Intentionality metrics: tools display metrics that evaluate the degree of human intention in the creative process (noting that these metrics are necessarily reductionist proxies and may become subject to Goodhart's/Campbell's law). For example, a generative design tool might analyse the uniqueness of user queries and the divergence of the output from standard templates. Visibilising the invisible effects of mechanised convergence may encourage users to engage more deeply with the work and make conscious, deliberate, individual choices.

    \item Participatory AI artefacts: artefacts are intentionally left incomplete by AI, requiring human participation\footnote{though not ``collaboration'' \citep{sarkar2023humanaicollaboration}} to finalise. For instance, a participatory tool generates the outline of a web design but leaves decisions about colour schemes and typography to the user. Conversely, a tool refuses to generate an outline, requiring the user to form a rough intention independently, before assisting by filling in details.
\end{itemize}

% **Ethical Hackathons:**
% Consider an event where programmers gather to tackle ethical challenges posed by AI, such as bias in machine learning models. These **Ethical Hackathons** encourage programmers to confront the anxiety of existential freedom by making choices that have far-reaching consequences. By developing solutions that prioritize fairness and transparency, programmers act in ways that respect the freedom of others, aligning with the existentialist ethos as advocated by thinkers like Simone de Beauvoir.

These speculations are not meant to be concrete proposals, but rather simply representative ideas of a future where the existentialist values of freedom, authenticity, and intentionality are preserved and enhanced through GenAI. They are limited in vision, representing only the lines of sight from where we stand today, and unable to anticipate the adjacent possibles of where we might travel. 

\section{Conclusion}

Programming is undeniably changing under the influence of GenAI. Intention appears to be all one needs to create software. But the notion that GenAI offers a neutral, unencumbered path to realising intentions is a mirage. Contrary to the assumption that GenAI merely executes human intentions, it also shapes them. At the very least, GenAI can induce ``mechanised convergence'', homogenising creative output, and reducing diversity in thought. There is therefore a risk of creating a ``stuck'' culture that recycles an old set of convergent ideas instead of fostering a new set of divergent ones.

In seeking a way through this problem we have encountered a variety of sources that we can draw upon to precipitate the active cultivation of intention: evolutionary pressures, the need for dignity and self-actualisation, constraints, subversion, materiality, and resistance. Finally, we discussed how the problem of intention resonates with the existentialist pursuits of freedom, identity, and authenticity. While this discussion of existentialism is necessarily cursory, limited, flawed, and provisional, its aim has been to situate the problems posed by GenAI to intentionality in the broadest possible scope.\footnote{\cite{Camus1955} describes existence (suicide) as the only truly serious philosophical problem.}

Programming must go beyond specification and embody the active cultivation of intentions. Existentialist philosophy offers a proactive, prescriptive framework for understanding the formation of human intentions as a process that ought to be held as deeply personal, ethically charged, and fundamentally free. It teaches us that to be human is to be involved in a continuous project of becoming. After all -- one is not born, but rather becomes, a programmer.

\section{Acknowledgements}
Thanks to Sean Rintel and Lev Tankelevitch for helping review drafts of this paper. I am especially grateful to Ava Scott and Richard Banks for their generous and helpful reflections.

% Thank you to Mariana M\unichar{259}r\unichar{259}\unichar{537}oiu for updating this template for PPIG 2016, to Luke Church and Alan Blackwell for defining this template and to Eleonora Bilotta, Thomas Green and Paola Kathuria for their help in defining and testing this template.

% \newpage

\bibliography{references}
\bibliographystyle{apacite} 
\end{document}